\documentclass[prletter,showpacs,twocolumn,typeset,superscriptaddress]{revtex4}   
\usepackage{amsfonts}
\usepackage{amsmath}
\usepackage{amssymb}
\usepackage{graphicx}
\usepackage{epsfig}
\usepackage{citesort}

\usepackage{dcolumn}
\usepackage{bm}

\begin{document}

\preprint{APS}

\title{Measurement-driven quantum evolution from a known state}
\author{Luis Roa}
\affiliation{Center for Quantum Optics and Quantum Information, Departamento de F\'{\i}sica,
Universidad de Concepci\'{o}n, Casilla 160-C, Concepci\'{o}n,
Chile.}
\author{G. A. Olivares-Renter\'{\i}a}
\affiliation{Center for Quantum Optics and Quantum Information, Departamento de F\'{\i}sica,
Universidad de Concepci\'{o}n, Casilla 160-C, Concepci\'{o}n,
Chile.}
\author{M. L. Ladr\'{o}n de Guevara}
\affiliation{Departamento de F\'{\i}sica, Universidad Cat\'{o}lica
del Norte, Casilla 1280, Antofagasta, Chile.}
\author{A. Delgado}
\affiliation{Center for Quantum Optics and Quantum Information, Departamento de F\'{\i}sica,
Universidad de Concepci\'{o}n, Casilla 160-C, Concepci\'{o}n,
Chile.}
\date{\today}

\begin{abstract}
We study the problem of driving a known initial quantum
state onto a known pure state without using a unitary evolution.
This
task can be achieved by means of von Neumann measurement processes,
introducing $N$ observables which are consecutively measured
in order to approach the state of the system to the target state.
We
proved that the probability of projecting onto the target state can be increased meaningfully
by adding suitable observables to the process, that is, it converges to $1$ when $N$ increases.
We also discuss a physical implementation of this scheme.

\end{abstract}

\pacs{03.67.-a, 03.65.-w}
\maketitle

\section{Introduction}

The problem of controlling quantum systems has been a renewed
subject of study. Quantum computing is basesed on the existence of a
set of universal quantum gates which, concatenated, allow one to
implement any unitary transformation within a fixed level of
accuracy. These quantum gates are implemented through the controlled
manipulation of the interactions among different physical
systems. Quantum communication protocols, such as quantum
teleportation \cite{Bennett}, entanglement swapping \cite{Zukowsky}
and dense coding, also require the precise application of some
unitary transformations in a finite set of transformations. A
related problem has also been addressed in the contex of quantum
control \cite{Schirmer}. There, the goal is to drive the evolution
of an initial, possibly mixed, state to a state having a
predetermined expectation value of some observable. This evolution
is also considered to be unitary.

In this article we study the control of quantum systems in the case
where it is not possible to resort to unitary transformations. Our
main goal is to map a \textit{known} quantum state onto another
known state via a sequence of measurements with the highest possible
success probability, that is, a controlled evolution via
measurements only. It has been shown \cite{Roa} that the mapping of
an \textit{unknown} quantum state onto a known pure state can be
optimally implemented with the help of two observables only. In this
case, the highest success probability is achieved when the
eigenstates of the two observables define mutually unbiased bases.
It has also been shown that, when the system subjected to the
measurements is affected by a decoherence mechanism, only one
observable is required \cite{Roa2}.

Here we study the case of driving by von Neumann processes \cite{vonN} a known initial state makeing use of more than two
observables. First we analyze the problem of two observables.
Thereafter, we show that a new observable can be added in order to
achieve a further increase in the success probability. By means of
numerical simulations we show that the success probability rapidly
approaches the unity when the number of observables increases.

\section{Driving the evolution by two observables}   \label{Sec1}

Let us start by supposing that a quantum system is in a \textit{known} $\rho$ state. Our
goal is to drive the system to the \textit{known} $|\zeta\rangle$ target state by measurements only.
If we measure the $\hat{\zeta}$ observable, whose eigenstates are $\left\{|\zeta\rangle,|\zeta_{\perp}\rangle\right\} $,
the probability of projecting to the $|\zeta\rangle$ target state is $p_{d}=\langle\zeta|\rho|\zeta\rangle$.
Natural questions arise: is it possible to increase this
direct probability $p_{d}$ by making use of an intermediate measurement of another
observable $\hat{\theta}$? And, if it is possible, then how is the relation
among $\rho$, $\hat{\zeta}$, and $\hat{\theta}$ which maximizes such probability?

So, in order to approach the state of the system to the $|\zeta\rangle$ target state \cite{comment},
we first measure an observable $\hat{\theta}$ which has the $\left\{|0_1\rangle,|1_1\rangle\right\}
$ eigenstates. As a second step we perform a measurement of $\hat{\zeta}$.
Thus,
the probability of reaching the $|\zeta\rangle$ target through one eigenstate of
$\hat{\theta}$ followed by a measurement of $\hat{\zeta}$ is given by
\begin{equation}
p_{1,s}=\langle0_1|\rho|0_1\rangle|\langle 0_1|\zeta\rangle|^{2}+\langle
1_1|\rho|1_1\rangle|\langle1_1|\zeta\rangle|^{2}.  \label{p1s}
\end{equation}
Making use of the normalization of $\rho$ and $|\zeta\rangle$, and of the orthonomalization of $|0_1\rangle$ and $|1_1\rangle$
the previous expression can be cast in the form
\begin{align}
p_{1,s} &  =\langle\zeta|\rho|\zeta\rangle-2\left\vert \langle 1_1|\zeta
\rangle\langle\zeta|0_1\rangle\right\vert ^{2}\left(
\langle\zeta|\rho|\zeta\rangle-\langle\zeta_{\perp}|\rho|\zeta_{\perp}
\rangle\right) \nonumber \\
&  +(2|\langle
0_1|\zeta\rangle|^{2}-1)\left[  \langle 0_1|\zeta\rangle\langle\zeta_{\perp}|0_1\rangle\langle
\zeta|\rho|\zeta_{\perp}\rangle+\text{c.c.}\right].\label{p1s2}
\end{align}
The second term at the r.h.s. of Eq. (\ref{p1s2}) contributes to
increase $p_{1,s}$ with respect to $p_{d}$ when
$\langle\zeta_{\perp}|\rho|\zeta_{\perp}\rangle$ is higher than
$\langle\zeta|\rho|\zeta\rangle$, otherwise it helps to decrease
$p_{1,s}$ with respect to $p_{d}$. Meanwhile, the third term plays a
role when the $\rho$ initial state has non-diagonal elements
different from zero in the $\hat{\zeta}$ representation. If
$\rho=I/2$, being $I$ the identity, then $p_{1s}=1/2$, so that
$p_{1,s}$ is independent of the choice of $\hat{\theta}$. If the
$\rho$ initial state is diagonal in the $\hat{\zeta}$ representation,
then, when $\langle\zeta
|\rho|\zeta\rangle<\langle\zeta_{\perp}|\rho|\zeta_{\perp}\rangle$, it
requires a $\hat{\theta}$ observable unbiased to $\hat{\zeta}$ in
order to optimize the process, whereas when
$\langle\zeta|\rho|\zeta\rangle
>\langle\zeta_{\perp}|\rho|\zeta_{\perp}\rangle$, there is not any $\hat{\theta
}$ observable which allows to increase $p_{1,s}$ over the value of
$p_d$.

The third term of the r.h.s. of Eq. (\ref{p1s2}) contributes
maximally to $p_{1,s}$ when $\arg(\langle\zeta|\rho|\zeta_{\perp}
\rangle\langle\zeta_{\perp}|0_1\rangle\langle 0_1|\zeta\rangle)=0$ and
$|\langle0_1|\zeta\rangle|^{2}\geq1/2$, or
$\arg(\langle\zeta|\rho|\zeta_{\perp}
\rangle\langle\zeta_{\perp}|0_1\rangle\langle 0_1|\zeta\rangle)=\pi$
and $|\langle0_1|\zeta \rangle|^{2}\leq1/2$. Since both cases are
symmetric with respect to $|\langle 0_1|\zeta\rangle|^{2}=1/2$, in
the following we consider only the latter.

\begin{figure} [t]
\includegraphics[angle=360,width=0.40\textwidth]{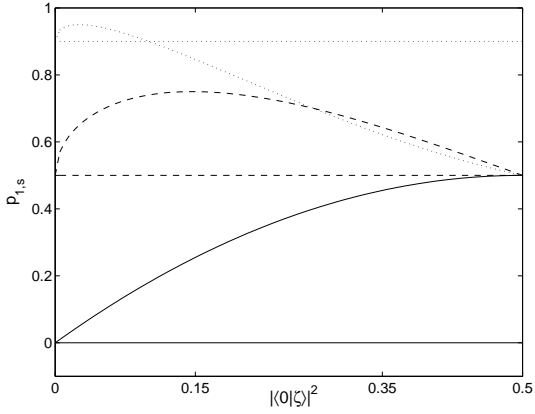}
\caption{Probability of success $p_{1,s}$ as a function of $\left\vert
\langle0_1|\zeta\rangle\right\vert ^{2}$ with $\langle\zeta|\rho|\zeta\rangle=0$
(solid), $\langle\zeta|\rho|\zeta\rangle=0.5$ (dash), and $\langle\zeta
|\rho|\zeta\rangle=0.9$ (dot). Horizontal lines are the respective $p_{d}$.} \label{figure1}
\end{figure}

Figure \ref{figure1} shows $p_{1,s}$ as a function of
$|\langle0_1|\zeta\rangle|^{2}$ for different initial values, say: $\langle\zeta|\rho|\zeta\rangle=0$
(solid line), $\langle\zeta|\rho|\zeta\rangle=0.5$ (dashed line), and
$\langle\zeta|\rho |\zeta\rangle=0.9$ (dotted line). In all of these
cases we have considered complete initial coherence, this is,
$|\langle\zeta|\rho|\zeta_{\perp}\rangle|
=\sqrt{\langle\zeta|\rho|\zeta\rangle\langle\zeta_{\perp}|\rho|\zeta_{\perp
}\rangle}$. The horizontal lines are the respective
$p_{d}=\langle\zeta |\rho|\zeta\rangle$.
We can see that for the considered initial conditions there is an
interval of $|\langle 0_1|\zeta\rangle|^{2}$ where $p_{1,s}$ is
higher than its associated $p_{d}$, and there is a particular value
of $|\langle0_1|\zeta\rangle|^{2}$ for which $p_{1,s}$ is maximum.

\begin{figure} [t]
\includegraphics[angle=360,width=0.40\textwidth]{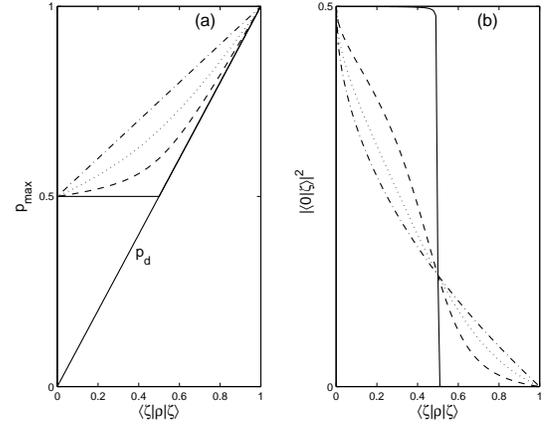}
\caption{(a) maximum probability of success $p_{max}$ as a function of
$\langle\zeta|\rho|\zeta\rangle$ for different $\gamma$ values:
$\gamma =1$ (dot-dash), $\gamma =0.7$ (dot),
$\gamma =0.4$ (dash), and $\gamma =0$ (solid).
(b) $|\langle0_1|\zeta\rangle|^{2}$ component as a function of
$\langle\zeta|\rho|\zeta\rangle$ for different $\gamma$ values:
$\gamma =1$ (dot-dash), $\gamma =0.7$ (dot),
$\gamma =0.4$ (dash), and $\gamma =0$ (solid).} \label{figure2}
\end{figure}

Let us examine what happen for a more general initial condition,
i.e., a $\rho$ initial state with $0\leq|
\langle\zeta|\rho|\zeta_{\perp}\rangle| \leq\sqrt{\langle
\zeta|\rho|\zeta\rangle\langle\zeta_{\perp}|\rho|\zeta_{\perp}\rangle}$,
where $\langle \zeta|\rho|\zeta\rangle\neq 1$. We look for conditions
under which the probability of success of the measurement process
$M(\zeta)M(\theta)$ is maximum. Optimizing Eq. (\ref{p1s2}) with
respect to $|\langle 0_1|\zeta\rangle|^2$, one finds that the maximum
value $p_{max}$ of $p_{1,s}$ is
\begin{equation}
p_{max}=\frac{\langle\zeta|\rho|\zeta\rangle}{2}
+\frac{1}{4}\left(
1+R\right),  \label{pmax}
\end{equation}
with
\begin{equation}
R=\sqrt{(1-\gamma^{2})(2\langle\zeta|\rho|\zeta\rangle-1)^{2}+\gamma^{2}},
\end{equation}
where we have defined the $\gamma$ coefficient by the equality
\begin{equation}
|\langle\zeta|\rho|\zeta_{\perp}\rangle|= \gamma\sqrt{\langle
\zeta|\rho|\zeta\rangle\langle\zeta_{\perp}|\rho|\zeta_{\perp}\rangle},
\hspace{0.1in}0\leq\gamma\leq1,\nonumber
\end{equation}
Fig. (\ref{figure2}.a) shows the maximum probability (\ref{pmax}) as a function
of $\langle\zeta|\rho|\zeta\rangle$ for different values of $\gamma$:
$\gamma=1$ (dot-dashed line), $\gamma=0.7$ (dotted line),
$\gamma=0.4$ (dashed line), and $\gamma=0$ (solid line). The
diagonal solid line corresponds to $p_d$. Notice that for all
$\gamma\neq0$ the optimal probability $p_{max}$ exceeds $p_d$  for
all values of $\langle\zeta|\rho|\zeta\rangle$. Also, larger values of
$\gamma$ result in larger values of $p_{max}$.

The $|0_1\rangle$ eigenstate of the $\hat{\theta}$
observable which optimizes $p_{1,s}$ has a component on the $|\zeta\rangle$ target
state given by
\begin{equation}
|\langle 0_1|\zeta\rangle|^{2}=\frac{1}{2}\left(  1-\frac
{1}{\sqrt{2}}\sqrt{1+\frac{2\langle\zeta|\rho|\zeta\rangle-1}{R}}\right).  \label{m}
\end{equation}
Fig. (\ref{figure2}.b) shows the square module (\ref{m}) of the $|0_1\rangle$ state
component onto the target state $|\zeta\rangle$ as a function of the initial probability of
the $|\zeta\rangle$ state for different $\gamma$ values.

When the initial state is pure ($\gamma=1$), $\rho=
|\psi\rangle\langle\psi|$, the square module of the $|0_1\rangle$ state
component on the target state $|\zeta\rangle$ becomes
\begin{equation}
|\langle 0_1|\zeta\rangle|^{2}=\frac{1-|\langle\psi|\zeta\rangle|}{2},
\label{hoo}
\end{equation}
which is a linear relation between a square module and a module of
two probability amplitudes. When the initial state is pure, $|\psi\rangle$, the average of (\ref{pmax}) on the Hilbert space reaches the value $3/4$.

Thus, we have found the $\hat{\theta}$ observable which
optimizes the fidelity or the probability
of taking the initial known state $\rho$ to the target
$|\zeta\rangle$ by means of von Neumann measurements only.

\section{Processing by {\textit N+1} observables}

First of all we study the case where three observables are implemented in order to achieve the target.
Hence the conlusions obtained are easily generalized when $N+1$ observables are considered.

Now we suppose that before measuring the observable $\hat{\zeta}$ we
measure two observables, say $\hat{\theta_1}$ followed by
$\hat{\theta_2}$, which define orthonormal bases
$\left\{|0_j\rangle,|1_j\rangle\right\}$, $j=1,2$, respectively. In
other words, we shall apply three consecutive von Neumann
measurement processes, first $M(\hat{\theta}_1)$ followed by
$M(\hat{\theta}_2)$ and finally by $M(\hat{\zeta})$, which shall be
denoted by the simple product
$M(\hat{\zeta})M(\hat{\theta}_2)M(\hat{\theta}_1)$. The probability
of driving the known initial state $\rho$ toward the $|\zeta\rangle$
target, by means of the von Neumann measurement
$M(\hat{\zeta})M(\hat{\theta}_2)M(\hat{\theta}_1)$ process, is given
by
\begin{eqnarray}
p_{2,s}&=&1-\langle 0_{1}|\rho |0_{1}\rangle -\left( 1-2\langle 0_{1}|\rho
|0_{1}\rangle \right)
\nonumber\\
&&\times
\left[ 1-\left\vert \langle 0_{2}|\zeta
\rangle \right\vert ^{2}-\left\vert \langle 0_{1}|0_{2}\rangle
\right\vert ^{2}\left( 1-2\left\vert \langle 0_{2}|\zeta \rangle
\right\vert ^{2}\right) \right],
\nonumber\\
\label{ps2}
\end{eqnarray}
where the quantities $\langle 0_{1}|\rho |0_{1}\rangle$ and
$|\langle 0_{1}|0_{2}\rangle|^{2}$ entering in $p_{2,s}$ are
considered to be functions of the coefficients of $\rho$ in the
basis of the $\zeta$ observable, of the quantities $|\langle
0_{1}|\zeta \rangle|^{2}$ and $|\langle 0_{2}|\zeta \rangle|^{2}$, and
of the phases $\varphi$ and $\phi$ of $\langle \zeta |\rho |\zeta
_{\perp }\rangle \langle 0_{1}|\zeta \rangle \langle \zeta _{\perp
}|0_{1}\rangle $ and $\langle 0_{1}|\zeta \rangle \langle \zeta
|0_{2}\rangle \langle 0_{1}|\zeta _{\perp }\rangle \langle \zeta
_{\perp }|0_{2}\rangle $, respectively.

The problem of optimizing the probability $p_{2,s}$, Eq.
(\ref{ps2}), leads to a set of nonlinear equations for the
quantities $\left\vert \langle 0_{1}|\zeta \rangle \right\vert ^{2}$,
$\left\vert \langle 0_{2}|\zeta \rangle \right\vert ^{2}$, $\varphi$
and $\phi$, which can not be analytically solved. However, we are
able to show that, under certain conditions, it is possible to
choose the observable $\theta_2$ in such a way that $p_{2,s}$
becomes higher than $p_{1,s}$.

The probability $p_{s,2}$ can be also written as
\begin{eqnarray}
p_{2,s} &=&p_{1,s}+\langle 0_{1}|\rho |0_{1}\rangle \left( \left\vert
\langle 0_{1}|0_{2}\rangle \right\vert ^{2}\left\vert \langle 0_{2}|\zeta
\rangle \right\vert ^{2}\right.  \nonumber \\
&&\left. +\left\vert \langle 0_{1}|1_{2}\rangle \right\vert ^{2}\left\vert
\langle 1_{2}|\zeta \rangle \right\vert ^{2}-\left\vert \langle 0_{1}|\zeta
\rangle \right\vert ^{2}\right)    \nonumber \\
&&+\langle 1_{1}|\rho |1_{1}\rangle \left( \left\vert \langle
1_{1}|0_{2}\rangle \right\vert ^{2}\left\vert \langle 0_{2}|\zeta \rangle
\right\vert ^{2}\right.    \nonumber \\
&&\left. +\left\vert \langle 1_{1}|1_{2}\rangle \right\vert ^{2}\left\vert
\langle 1_{2}|\zeta \rangle \right\vert ^{2}-\left\vert \langle 1_{1}|\zeta
\rangle \right\vert ^{2}\right),
\end{eqnarray}
where $p_{1,s}$ is given by Eq. (\ref{p1s}). Hence $p_{2,s}$ is higher than $p_{1,s}$ under the conditions:
\begin{equation}
\langle 0_{1}|\rho |0_{1}\rangle >\langle 1_{1}|\rho |1_{1}\rangle,  \label{c11}
\end{equation}
and
\begin{equation}
\left\vert \langle 0_{1}|0_{2}\rangle \right\vert ^{2}\left\vert \langle
0_{2}|\zeta \rangle \right\vert ^{2}+\left\vert \langle 0_{1}|1_{2}\rangle
\right\vert ^{2}\left\vert \langle 1_{2}|\zeta \rangle \right\vert
^{2}>\left\vert \langle 0_{1}|\zeta \rangle \right\vert ^{2}.  \label{c12}
\end{equation}
The condition (\ref{c12}) means that the basis
$\{|0_2\rangle,|1_2\rangle\}$ has to be chosen in a way such that
the probability of taking the state $|0_1\rangle$ to the state
$|\zeta\rangle$ by means of the $M(\zeta)M(\theta_2)$ process be
higher than the probability of taking the state $|0_1\rangle$ to the
state $|\zeta\rangle$ by means of the $M(\zeta)$ process. We have
already shown, in section \ref{Sec1}, that such a choice is always possible. The probability
$p_{2,s}$ is higher than $p_{1,s}$ also under the conditions:
\begin{equation}
\langle 0_{1}|\rho |0_{1}\rangle <\langle 1_{1}|\rho |1_{1}\rangle,
\end{equation}
and
\begin{equation}
\left\vert \langle 1_{1}|0_{2}\rangle \right\vert ^{2}\left\vert \langle
0_{2}|\zeta \rangle \right\vert ^{2}+\left\vert \langle 1_{1}|1_{2}\rangle
\right\vert ^{2}\left\vert \langle 1_{2}|\zeta \rangle \right\vert
^{2}>\left\vert \langle 1_{1}|\zeta \rangle \right\vert ^{2}.
  \label{c22}
\end{equation}
The latter condition has the same meaning as the (\ref{c12})
inequality, but in this case starting from the $|1_1\rangle$ state
instead of from the $|0_1\rangle$ state. This condition can also be
always satisfied.

The above result can be generalized to the case of $N$ observables $theta_i$.
In this case we suppose that, before measuring the observable $\hat{\zeta}$,
we measure $N$ observables, say
$\hat{\theta_1}$, $\hat{\theta_2}$, $\ldots$, $\hat{\theta}_N$, each one
defining an orthonormal basis $\left\{|i_j\rangle\right\}$ respectively,
with $i_j=0,1$ and $j=1,2\ldots,N$.
The probability of driving the
known initial state $\rho$ towards the $|\zeta\rangle$ target,
by means of the von Neumann measurement processes
$M(\hat{\zeta})M(\hat{\theta}_N)\ldots M(\hat{\theta}_2)M(\hat{\theta}_1)$, can be calculated recursively as
\begin{equation}
p_{N,s}=\langle \zeta |\rho _{N}|\zeta \rangle,
\label{pn}
\end{equation}
where $\rho _{N}$ is given by
\begin{equation}
\rho _{N}=\sum_{i_{N}=0}^{1}\langle i_{N}|\rho _{N-1}|i_{N}\rangle
|i_{N}\rangle\langle i_{N}|.
\end{equation}
The difference $\Delta=p_{N+1,s}-p_{N,s}$ can be read as
\begin{eqnarray}
\Delta&=&\sum_{i_{N}=0}^{1}\langle i_{N}|\rho _{N-1}|i_{N}\rangle \nonumber             \\
&\times &(\sum_{i_{N+1}=0}^{1}|\langle i_{N}|i_{N+1}\rangle|^2           
|\langle i_{N+1}|\zeta \rangle|^2-|\langle i_{N}|\zeta \rangle|^2).
\nonumber\\
\label{nc}
\end{eqnarray}
The positivity of this difference is guaranteed under the conditions
\begin{equation}
\langle 0_{N}|\rho _{N-1}|0_{N}\rangle >\langle 1_{N}|\rho
_{N-1}|1_{N}\rangle,
\end{equation}
and
\begin{equation}
\sum_{i_{N+1}=0}^{1}|\langle 0_{N}|i_{N+1}\rangle |^2|\langle i_{N+1} | \zeta\rangle |^2
>|\langle0_{N} |\zeta\rangle |^2.
\label{c01at}
\end{equation}
The latter condition means that the basis $\{|0_{N+1}\rangle,|1_{N+1}\rangle\}$ must be
chosen in a way such that the probability of taking state $|0_N\rangle$ to state
$|\zeta\rangle$ by means of the $M(\zeta)M(\theta_{N+1})$ process be higher than the
probability obtained by means of the $M(\zeta)$ process.
In section \ref{Sec1} we have already shown that this can be always achieved.
The positivity of Eq. (\ref{nc}) is also satisfied if
$
\langle 0_{N}|\rho _{N-1}|0_{N}\rangle <\langle 1_{N}|\rho
_{N-1}|1_{N}\rangle
$
and
\begin{equation}
\sum_{i_{N+1}=0}^{1}|\langle 1_{N}|i_{N+1}\rangle |^{2}|\langle i_{N+1}|\zeta
\rangle |^{2}>|\langle 1_{N}|\zeta \rangle |^{2},
\end{equation}
which has the same meaning as the (\ref{c01at}) condition,
starting from the $|1_{N+1}\rangle$ state instead of from $|0_{N+1}\rangle$.

\begin{figure} [t]
\includegraphics[angle=360,width=0.40\textwidth]{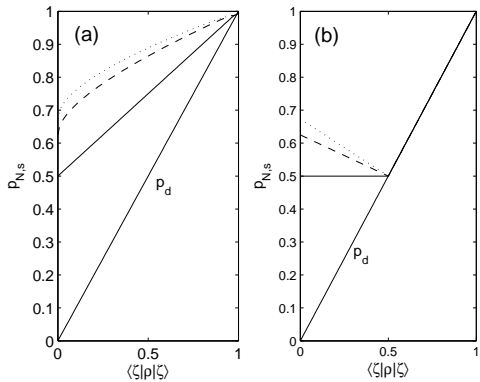}
\caption{Maximum value of $p_{N,s}$ probability as a function of the
$\langle\zeta|\rho|\zeta\rangle$ initial probability for different $N$ values:
$N=1$ (solid), $N=2$ (dash), $N=3$ (dot), for: (a) an initial pure state $\gamma=1$ and (b) a mixed initial state $\gamma=0$. The diagonal solid line corresponds to $p_d$.
} \label{figure3}
\end{figure}

Thus, we have shown that the probability of success can be increased
by adding suitable observables to the process. Since each suitable
$\theta_i$ observable depends on the $\rho$ initial state and the
$|\zeta\rangle$ target, we can conclude that as $N$ goes to infinity,
the fidelity and the probability of finding $|\zeta\rangle$ will go
to $1$. This conclusion also can be obtained by studying the Hilbert-Schmidt distance \cite{Lee}
between $\rho_N$ and $|\zeta\rangle\langle\zeta|$.

Figure \ref{figure3} shows the results of a numerical
simulation which finds the (\ref{pn}) probability for a given set of bases $\left\{|0_j\rangle,|1_j\rangle\right\}$ ($j=1,2,\dots N$) with $N$ and $\rho$ fixed.
On it is plotted the maximum value of $p_{N,s}$ as a function of the
$\langle\zeta|\rho|\zeta\rangle$ initial probability for different $N$ values:
$N=1$ (solid), $N=2$ (dash), $N=3$ (dot), when (a) $\gamma=1$ and (b) $\gamma=0$.
In Fig. \ref{figure3}.a, which corresponds an initial pure state, we can see that $p_{N,s}$ increases with respect
to $p_d$ (the diagonal) for all $\langle\zeta|\rho|\zeta\rangle$ initial probability.
In Fig. \ref{figure3}.b, which corresponds to an initial mixed state (diagonal in the $\left\{|\zeta\rangle,|\zeta_\perp\rangle\right\}$) basis),
we can see that $p_{N,s}$ increases with respect to $p_d$ (the diagonal) only for $\langle\zeta|\rho|\zeta\rangle < 1/2$.

\section{Conclusions}

In summary, we have studied the problem of driving a known initial
quantum state onto a known pure state without using any unitary
transformation. This task can be achieved by means of von Neumann
measurement processes, introducing $N$ observables which are
consecutively measured in order to get the state closer to the target
state. We proved that the probability of projecting onto the target
can be increased by adding suitable observables to the process.
Since each of these suitable observable depends on the $\rho$
initial state and on the $|\zeta\rangle$ target, we conclude that as
$N$ increase the probability of finding $|\zeta\rangle$ goes
to $1$.

For a physical implementation of the above described process one
could address the problem of keeping the initial flux of a beam composed of a
collection of systems in the same state, each one exposed to a
postselection-measurement procedure. For instance, let us consider a
source of monochromatic and vertically linear polarized photons
\cite{Peres}. In order to obtain photons in a horizontally linear
polarized state it is required to put a linear polarizer in their
path. Implementing two linear polarizers in a suitable
configuration, the outcome flux with horizontal linear polarization
is decreased fifty per cent with respect to the incoming flux. By
implementing more than two linear polarizers, as is suggested above,
the output flux of the beam can be increased meaningfully and it can be approached to
the initial flux, depending on the number of linear polarizers
arranged suitably. Since in this scheme only one component of the
each linear polarized flux contributes to the success probability, it
will converge a little more slowly than our protocol, however it will also go to $1$ as the number of linear
polarizers arranged suitably increases, preserving approximately the initial flux. A nonlinear crystal can
change the polarization of a photon while preserving the flux;
however, it also preserves the initial mix degree. In our scheme,
independently of the initial mix degree, the output is pure.

Further studies could be generalized considering a $d$-dimensional
Hilbert space.

This work was supported by Milenio Grant ICM P02-49F, and FONDECYT
Grants 1030671, 1040591, and 1040385.


\begin{thebibliography}{99}
\bibitem{Nielsen} M. A. Nielsen and I. L. Chuang, {\it Quantum Computation and Quantum Information} (Cambridge University Press, Cambridge, U.K., 2000); G. Alber et al. {\it Quantum Information} (Springer, Berlin, 2001).
\bibitem{Landauer} R. Landauer, Phys. Lett. A {\bf 217}, 188 (1996).
\bibitem{Bennett} C. H. Bennett, G. Brassard, C. Cr\'{e}peau, R. Jozsa, A. Peres, and W. K. Wootters, Phys. Rev. Lett. \textbf{70}, 1895 (1993).
\bibitem{Zukowsky} M. $\dot{\text Z}$ukowsky, A. Zeilinger, M. A. Horne, and A. K. Ekert, Phys. Rev. Lett. \textbf{71}, 4287 (1993).
\bibitem{Wootters} W. K. Wootters and W. H. Zurek, Nature \textbf{299}, 802 (1982).
\bibitem{Duan} L. M. Duan and G. C. Guo, Phys. Rev. Lett. \textbf{80}, 4999 (1998).
\bibitem{Pati} A. K. Pati and S. L. Braunstein, Nature (London) \textbf{404}, 164 (2000).
\bibitem{Zhao1} Z. Zhao, Y. A. Chen, A. N. Zhang, T. Yang, H. J. Briegel, and J. W. Pan, Nature (London) \textbf{430}, 54 (2004).
\bibitem{Pan1} J. W. Pan, M. Daniell, S. Gasparoni, G. Weihs, and A. Zeilinger, Phys. Rev. Lett. \textbf{86}, 4435 (2001).
\bibitem{Pan2} J. W. Pan, D. Bouwmeester, M. Daniell, H. Weinfurter, and A. Zeilinger, Nature (London) \textbf{403}, 515 (2000).
\bibitem{Bouwmeester1} D. Bouwmeester, J. W. Pan, M. Daniell, H. Weinfurter, and A. Zeilinger, Phys. Rev. Lett. \textbf{82}, 1345 (1999).
\bibitem{Boschi} D. Boschi, S. Branca, F. De Martini, L. Hardy, and S. Popescu, Phys. Rev. Lett. \textbf{80}, 1121 (1998).
\bibitem{Bouwmeester2} D. Bouwmeester, J. W. Pan, K. Mattle, M. Eibl, H. Weinfurter, and A. Zeilinger, Nature (London) \textbf{390}, 575 (1997).
\bibitem{Lamas} A. Lamas-Linares, C. Simon, J. C. Howell, and D. Bouwmeester, Science \textbf{296}, 712 (2002).
\bibitem{Martini} F. De Martini, V. Bu$\check{\text{z}}$ek, F. Sciarrino, and C. Sias, Nature (London) \textbf{419}, 815 (2002).
\bibitem{Ziman} M. Ziman, P. $\check{\text{S}}$telmachovi$\check{\text{c}}$, V. Bu$\check{\text{z}}$ek, M. Hillery, M. Scarani, and N. Gisin, Phys. Rev. A \textbf{65}, 042105 (2002).
\bibitem{Barenco} A. Barenco, C. H. Bennett, R. Cleve, D. P. DiVincenzo, N. Margolus, P. Shor, T. Sleator, J. A. Smolin, and H. Weinfurter, Phys. Rev. A \textbf{52}, 3457 (1995).
\bibitem{Schirmer} S. G. Schirmer and J. V. Leahy, Phys. Rev. A \textbf{63}, 025403 (2001); F. Albertini and D. D. Alessandro, quant-ph/0106128.
\bibitem{Roa} L. Roa, A. Delgado, M. L. Ladr\'{o}n de Guevara, and A. B. Klimov, Phys. Rev. A \textbf{73}, 012322 (2006).
\bibitem{Roa2} Luis Roa and G. A. Olivares-Renter\'{\i}a, Phys. Rev. A. \textbf{73}, 062327 (2006).
\bibitem{vonN} J. von Neumann, Ann. Math. \textbf{32}, 191 (1931); J. von Neumann, {\it Mathematische Grundlagen der Quantenmechanik} (Springer, Berlin, 1932).
\bibitem{comment} \textquotedblleft To approach the state of the system to the $|\zeta\rangle$ target state\textquotedblright can be understood in the sense that the fidelity or probability to obtain the $|\zeta\rangle$ state goes closer to $1$.
\bibitem{Lee} J. Lee, M. S. Kim, and $\check{\text{C}}$. Brukner, Phys. Rev. Lett. \textbf{91}, 087902 (2003).
\bibitem{Peres} Asher Peres, {\it Quantum Theory: Concepts and Methods} (Kluwer Academic Publishers, 1998).
\end{thebibliography}
\end{document}